\documentclass[showpacs,preprintnumbers,amsmath,amssymb]{revtex4}

\usepackage[dvips]{color}
\definecolor{Red}{rgb}{1.00, 0.00, 0.00}

\usepackage{amsmath,amssymb}
\usepackage{graphicx}
\usepackage{dcolumn}
\usepackage{bm}
\usepackage{epstopdf}

\newcommand{\al}{\alpha}
\newcommand{\be}{\beta}
\newcommand{\ga}{\gamma}
\newcommand{\Ga}{\Gamma}
\newcommand{\de}{\delta}
\newcommand{\De}{\Delta}
\newcommand{\ep}{\varepsilon}

\newcommand{\om}{\omega}

\newcommand{\p}{\partial}
 
\newcommand{\txt}{\textstyle}
\newcommand{\dsp}{\displaystyle}
\newcommand\eqn[1]{(\ref{#1})}      
\newcommand{\beq}{\begin{equation}}
\newcommand{\eeq}{\end{equation}}
\newcommand{\ba}{\begin{array}}
\newcommand{\ea}{\end{array}}
\newcommand{\bea}{\begin{eqnarray}}
\newcommand{\eea}{\end{eqnarray}}
\newcommand{\bi}{\begin{itemize}}  
\newcommand{\ei}{\end{itemize}}
\newcommand{\ben}{\begin{enumerate}} 
\newcommand{\een}{\end{enumerate}}

\newcommand{\half} {{\txt \frac{1}{2}}}

\newcommand{\ee}[1]{\times 10^{#1}}

\newcommand{\s}{{\rm s}}
\newcommand{\g}{{\rm g}}

\newcommand{\cm}{{\rm cm}}

\newcommand{\MeV}{{\rm MeV}}

\renewcommand{\a}{a}  

\usepackage[normalem]{ulem}  

\newcommand{\feyn}[1]{
  \setbox0=\hbox{\ensuremath{#1}}
  \hbox to\wd0{\hbox to0pt{\hbox to\wd0{\hss/\hss}\hss}\box0}}

\newcommand{\ebar}{\bar{e}}
\newcommand{\mubar}{\bar{\mu}}
\newcommand{\nue}{\nu_e}
\newcommand{\numu}{\nu_\mu}
\newcommand{\nuebar}{{\bar{\nu}}_e}
\newcommand{\numubar}{{\bar{\nu}}_\mu}
\newcommand{\gaeff}{{\gamma_{\rm eff}}}

\newcommand{\vp}{{\bf p}}

\newcommand{\projl}{\left(1\!-\!\gamma^5\right)} 
\newcommand{\barC}{\bar{C}}
\newcommand{\qmin}{q_{\rm min}}

\begin{document}

\preprint{}

\title{
Leptonic contribution to the bulk viscosity of nuclear matter
}

\author{Mark G. Alford and Gerald Good}%
\affiliation{Physics Department, Washington University,
St.~Louis, MO~63130, USA}

\date{10 Nov 2010} 

\begin{abstract}
For $\be$-equilibrated nuclear matter 
we estimate the contribution to 
the bulk viscosity from purely
leptonic processes, namely the conversion
of electrons to and from muons.
For oscillation frequencies in the kHz range, we find that this process
provides the dominant contribution to the bulk viscosity when the
temperature is well below the critical temperature for
superconductivity or superfluidity of the nuclear matter.
\end{abstract}

\pacs{26.60.-c,21.65.Cd,67.10.Jn}

\maketitle

\section{Introduction}
\label{sec:intro}

The bulk viscosity of nuclear matter plays an important role
in the damping of oscillations in neutron stars.
One well-known example is ${\it r}$-modes, which, if
the interior of the star is a perfect (dissipationless) fluid,
become unstable with respect to the emission
of gravitational waves \cite{Andersson:1997xt,Lindblom:2000jw,Owen:1998xg}. 
This emission acts as a brake on the rotation of the star. However,
$r$-mode spindown will not occur if the $r$-mode is sufficiently
strongly damped, for example by shear or bulk viscosity of the
matter in the interior of the star.
It is therefore important to calculate of the
bulk viscosity of the various candidate phases in a neutron star.
Several calculations exist in the literature, for nuclear
\cite{Haensel:2000vz,Haensel:2001mw-ADS,Andersson:2004aa,Chatterjee:2007qs,Gusakov:2007px,Chatterjee:2007ka} and hyperonic
\cite{Lindblom:2001hd,Haensel:2001em-ADS,Chatterjee:2007iw-ADS} 
as well as for unpaired quark
matter \cite{Madsen:1999ci,Madsen:1992sx,astro-ph/0703016} and various
color-superconducting phases
\cite{Manuel:2004iv,Alford:2006gy,astro-ph/0607643,Manuel:2007pz,Dong:2007ax,Alford:2007rw}.

In this paper we will study $\be$-equilibrated nuclear matter.
We define the chemical potential for charged leptons to be
$\mu_l=-\phi/e$ where $\phi$ is the electrostatic potential
and $e$ is the positron charge.
We will assume that the density is high enough that
$\mu_l$ is greater than the mass of the muon,
so the matter consists of neutrons, protons, electrons and muons.
Such matter is expected to exist in the core of the star.
In previous calculations of bulk viscosity of $npe\mu$ nuclear matter
the focus has been on the contribution from interconversion
of neutrons and protons via weak interactions. But nuclear matter
at neutron-star densities is expected to show Cooper pairing of
protons (superconductivity) or neutrons (superfluidity)
\cite{Dean:2002zx,Muther:2005cj,Gandolfi:2009tv} either of which
will suppress interconversion by a factor 
of order $\exp(-\De/T)$, where $\De$
is the energy gap at the Fermi surface and $T$ is the temperature.
This opens up the possibility that, in superfluid or superconducting
phases, the dominant contribution
to the bulk viscosity might come from purely leptonic processes.
The relevant process is conversion of electrons to
muons (and vice versa) via either the direct Urca process or the modified
Urca process. The direct Urca
leptonic conversion process is forbidden by
energy and momentum conservation: in converting
an electron
near its Fermi surface to a muon near its Fermi surface, the
change in free energy is very small (of order $T$), so the
emitted neutrinos carry momentum and energy of this order;
but the change of momentum of the charged lepton is large, at least
$\qmin =\, \mu_l - \sqrt{\mu_l^2-m_\mu^2}$, and the low-energy neutrino
cannot carry this much momentum.
However, the modified Urca process can
occur; for example, two electrons with energy slightly above the Fermi
energy can scatter to an electron and a muon with energies near the
Fermi energy, or an electron and muon can scatter to two muons. 
The strongest interaction between leptons is electromagnetism, so
this process proceeds via exchange of a photon, whose propagator
should include the effects of screening by the nuclear medium. As the
temperature decreases, the process will become suppressed as the Fermi
distributions assume their zero-temperature step function profiles,
but at finite temperature the modified Urca process will result in a
non-zero contribution to the bulk viscosity.

We calculate the leptonic bulk viscosity arising
from the processes $e + \ell \rightleftharpoons \mu +
\ell + \nu + \bar{\nu}$, where $\ell = e$ or $\mu$.
All our calculations are in the ``subthermal'' regime where
the density oscillation has a small amplitude, and the bulk viscosity
is independent of that amplitude.
We conclude that, if the protons and neutrons are both ungapped,
i.e if there is neither superfluidity nor superconductivity,
then the bulk viscosity from these
purely leptonic processes is several orders of magnitude smaller than
that from the nucleonic processes.
However, once the temperature drops below the critical value
for Cooper pairing of the protons or neutrons, the nucleonic
bulk viscosity at frequencies $\gtrsim 10$\,Hz
is strongly suppressed, and leptonic processes become
the dominant source of bulk viscosity at those frequencies.

In section \ref{sec:leptons}, we lay out the process for calculating
the bulk viscosity of a two-component leptonic system under
application of a periodic volume and pressure perturbation. A crucial
component of this calculation is the conversion rate between electrons
and muons, which is discussed in \ref{sec:rate}. In section \ref{sec:results} we
show the numerical results of our calculations and how they compare to
the bulk viscosity resulting from modified Urca equilibration of
the nucleon population.

\section{Bulk viscosity of leptons}
\label{sec:leptons}

First we write down a general expression for bulk viscosity
in a two-species system, arising from interconversion of the
two species. Then we specialize to the case of electrons and
muons in nuclear matter.

\subsection{Bulk viscosity of a two-species system}
\label{sec:two_species}

We assume that the system experiences a 
small-amplitude driving oscillation
\beq
\ba{rcl}
V(t) &=& \bar{V} + {\rm Re}(\de V\, e^{i\om t}) \\[1ex]
p(t) &=& \bar{p} + {\rm Re}(\de p\, e^{i\om t})
\ea
\label{epsilons}
\eeq
where the volume amplitude $\de V \ll \bar V$
is real by convention, and the resultant
pressure oscillation $p(t)$ is complex.
The average power dissipated per unit volume is
\beq
\frac{dE}{dt} = -\frac{1}{\tau \bar V}\int_0^\tau p(t)\frac{dV}{dt}dt
 =  -\frac{1}{2}\dsp \om \,{\rm Im}(\de p)\,\frac{\de V}{\bar V} \ ,
\label{dEdt}
\eeq
where $\tau=2\pi/\om$, so the bulk viscosity is \cite{Madsen:1992sx}
\begin{equation}
\zeta = 
\frac{2 \bar V^2}{\om^2 (\de V)^2} \frac{dE}{dt} 
 = -\frac{{\rm Im}(\de p)}{\de V} \frac{\bar V}{\om} \ .
\label{zeta_def}
\end{equation}
We will determine ${\rm Im}(\de p)$, which will be negative.  We will
assume that heat arising from dissipation is conducted away quickly,
so the whole calculation is performed at constant temperature $T$.  We
assume that our system contains two particle species $e$ and $\mu$, and
the state of the system is determined by the corresponding chemical
potentials $\mu_e$ and $\mu_\mu$.  The total number of electrons
and muons is conserved, and equilibrium is established via the
conversion process $e\leftrightarrow\mu$.  For simplicity of
presentation and of the final expressions, it is better to work in
terms of charged lepton number $l$ and electron-muon asymmetry $a$, so
pressure is a function of $\mu_l$ and $\mu_\a$, where
\beq
\ba{rcl@{\qquad}rcl}
\mu_l &=& \half(\mu_e + \mu_\mu)  & n_l &=& \dsp n_e + n_\mu 
  = \left.\frac{\p p}{\p\mu_l}\right|_{\mu_\a}\\[3ex]
\mu_\a &=& \half(\mu_e - \mu_\mu)  & n_\a &=& \dsp n_e - n_\mu 
  = \left. \frac{\p p}{\p\mu_\a}\right|_{\mu_l}
\ea
\label{mudef}
\eeq
From now on all partial derivatives with respect to $\mu_l$
will be assumed to be at constant $\mu_\a$, and vice versa.
In beta-equilibrium, $\mu_\a$ is zero.
The variations in the chemical potentials are expressed in terms
of complex amplitudes $\de\mu_l$, and $\de\mu_\a$,
\beq
\ba{rcl}
\mu_l(t) &=& \bar{\mu}_l + {\rm Re}(\de \mu_l \, e^{i\om t}) \ , \\
\mu_\a(t) &=&\phantom{\bar{\mu}_l\, +\,} {\rm Re}(\de\mu_\a e^{i\om t}) \ .
\ea
\label{mu_epsilons}
\eeq
The pressure amplitude is then
\beq
\de p =
  \frac{\p p}{\p \mu_l}\Bigr|_{\mu_\a} \de \mu_l
 +\frac{\p p}{\p \mu_\a}\Bigr|_{\mu_l} \de \mu_\a
 = n_l \de \mu_l + n_\a  \de \mu_\a\ ,
\label{dp_full}
\eeq
From \eqn{dp_full} and \eqn{zeta_def} we find
\beq
\zeta = -\frac{1}{\om}\frac{\bar V}{\de V}\Bigl(
 \bar n_l{\rm Im}(\de\mu_l) + \bar n_\a{\rm Im}(\de\mu_\a ) \Bigr) \ .
\label{zeta}
\eeq
To obtain the imaginary parts of the chemical potential amplitudes,
we write down the rate of change of the corresponding conserved quantities,
\beq
\ba{rclcl}
\dsp \frac{dn_l}{dt} 
 &=&\dsp \frac{\p n_l}{\p \mu_l}\frac{d\mu_l}{dt}
        +\frac{\p n_l}{\p \mu_\a}\frac{d\mu_\a}{dt}
 &=&\dsp -\frac{n_l}{\bar V} \frac{dV}{dt} \ , \\[2ex]
\dsp \frac{dn_\a}{dt} 
 &=&\dsp \frac{\p n_\a}{\p\mu_l}\frac{\p \mu_l}{dt}
       + \frac{\p n_\a}{\p \mu_\a}\frac{d\mu_\a}{dt}
 &=&\dsp -\frac{n_\a}{\bar V} \frac{dV}{dt} - \Ga_{e\to\mu}^{\rm total} \ .
\ea
\label{ndots1}
\eeq
All the partial derivatives are evaluated at equilibrium, $\mu_l=\bar\mu_l$ 
and $\mu_\a=0$.
The right hand term on the first line expresses the fact that 
charge is conserved, so when a volume is compressed, the density
of charged leptons rises. 
On the second line, there is such a term from the
compression of the existing population of particles, but there is also a
rate of conversion $\Gamma_{e\to\mu}^{\rm total}$ of electrons to
muons, which reflects the fact that weak interactions will push the
lepton densities towards their equilibrium value.  For small
deviations from equilibrium we expect $\Ga_{e\to\mu}^{\rm total}$ to
be linear in $\mu_\a$, so it is convenient to write the rate in terms
of an average width $\ga_\a$, which is defined in terms of the total
rate by writing
\beq
\Gamma_{e\to\mu}^{\rm total}
= \ga_\a\,\frac{\p n_\a}{\p \mu_\a}\mu_\a\ .
\label{gamma_phi}
\eeq

We now substitute the assumed oscillations \eqn{epsilons} and
\eqn{mu_epsilons} in to \eqn{ndots1}, and solve to obtain
the amplitudes $\de\mu_l$ and $\de\mu_\a$ in terms of the amplitude $\de V$
and frequency $\om$ of the driving oscillation. Inserting their
imaginary parts in \eqn{zeta} we obtain the bulk viscosity, which is
conveniently expressed in terms of the susceptibilities
\beq
\ba{rcl}
\chi_{ll} &=&\dsp \frac{\p n_l}{\p \mu_l} \ ,\\[3ex]
\chi_{l\a} &=&\dsp \frac{\p n_l}{\p \mu_\a} = \frac{\p n_\a}{\p \mu_l}\ ,\\[3ex]
\chi_{\a\a} &=&\dsp \frac{\p n_\a}{\p \mu_\a} \ ,
\ea
\label{susceptibilities}
\eeq
all evaluated at equilibrium, $\mu_l=\bar\mu_l$, $\mu_a=0$.
Note that $\chi_{\a l}$ is the same as $\chi_{l\a}$ from \eqn{mudef}.
Defining 
\beq
\ba{rcl}
\gaeff &=&\dsp 
  \frac{ \chi_{ll}\chi_{\a\a}}{\chi_{ll}\chi_{\a\a}-\chi_{l\a}^2} \ga_\a  
= \frac{ \chi_{ll}}{\chi_{ll}\chi_{\a\a}-\chi_{l\a}^2} 
  \frac{\p \Gamma_{e\to\mu}^{\rm total}}{\p \mu_\a}\Bigr|_{\mu_\a=0},\\[4ex]
C &=&\dsp 
\frac{ (\chi_{ll} n_\a - \chi_{l\a} n_l)^2}{
  \chi_{ll}(\chi_{ll}\chi_{\a\a}-\chi_{l\a}^2)}\ ,
\ea
\label{C_gamma}
\eeq
we obtain the final result for the bulk viscosity 
in a two-species system,
\beq
\zeta = C\frac{ \ga_{\rm eff} }{\om^2 + \ga_{\rm eff}^2} \ .
\label{zeta_phi}
\eeq

From \eqn{zeta_phi} we can already see how the bulk viscosity 
of a two-species system depends
on the frequency $\om$ of the oscillation and the effective
equilibration rate $\ga_{\rm eff}$. 

At fixed equilibration rate, the bulk viscosity decreases
monotonically as the
oscillation frequency rises; it is roughly constant for
$\om\lesssim\ga_{\rm eff}$, and then drops off quickly as $1/\om^2$ for
$\om\gg\ga_{\rm eff}$.  

At fixed oscillation frequency $\om$, the bulk viscosity is a non-monotonic
function of the rate $\ga_{\rm eff}$. It is peaked at
$\ga_{\rm eff}=\om$, with a value 
\beq
\zeta_{\rm max} = \half C/\om \ .
\label{zeta_max}
\eeq
For $\ga_{\rm eff}\ll\om$ or $\ga_{\rm eff}\gg\om$
the bulk viscosity tends to zero.  Thus very fast and very slow
processes are not an important source of bulk viscosity.
As we will see below, for leptons in nuclear matter the
equilibration rate is sensitive to temperature but
the coefficient $C$ is not, so we expect $\zeta(T)$ to
be peaked at $\ga_{\rm eff}(T)=\om$, where the
oscillation frequency $\om$ is of order kHz for typical
oscillation modes of neutron stars.

\subsection{Leptons in nuclear matter}

In nuclear matter the leptonic chemical potential $\mu_l=\mu_e=\mu_\mu$
is much greater than the temperature and the electron mass, so we
can evaluate the susceptibilities \eqn{susceptibilities} at $m_e=T=0$.
Temperature dependence will come in only via the equilibration
rate $\ga_\a$.
Treating the electrons and muons as free fermions, we find
\beq
\ba{rcl}
\ga_{\rm eff} &=&\dsp \ga_a 
  \frac{(\mu_l+k_F)^2}{4\mu_l k_F}\ ,\\[3ex]
C &=& \dsp \frac{1}{9\pi^2}  m_\mu^2 k_F (\mu_l-k_F) \ .
\ea
\label{coeffs}
\eeq
where the muon Fermi momentum is given by $k_F^2 = \mu_l^2-m_\mu^2$.
Note that the bulk viscosity goes to zero as $m_\mu\to 0$ 
($m_\mu\to m_e$, really). This is because if the muons and
electrons have equal mass then under compression their relative densities
do not change, and there is no need for any equilibrating process, 
so the pressure is always in phase with the volume and no
dissipation occurs.

Even without calculating the rate of lepton number equilibration, we can now
estimate the amount of bulk viscosity that could possibly
arise from leptons. If the equilibrating weak interaction at
some temperature happened to have a rate that matched the typical oscillation
frequency of the star, $\om\approx 2\pi\times 1000$~Hz, 
and the lepton chemical potential had a relatively moderate
value of about $120\,\MeV$, we would obtain
from \eqn{zeta_max},
$\zeta_{\rm max} = 5.5\ee{22}~\MeV^3 = 7.5\ee{27}~\g\,\s^{-1}\cm^{-1}$.
This is at the upper end of typical nuclear bulk viscosities which
range up to $10^{28}\g\,\s^{-1}\cm^{-1}$ \cite{Haensel:2000vz}.
This motivates us to proceed with the calculation of the
rate of conversion of muons to and from electrons
via the weak interaction.

\section{Muon-electron conversion rate}
\label{sec:rate}

The muon-electron conversion rate $\Gamma_{e\to\mu}^{\rm total}$
consists of two partial rates,
\beq
\Gamma_{e\to\mu}^{\rm total} = \Gamma_{ee\to e\mu}^{\rm total} + \Gamma_{e\mu\to\mu\mu}^{\rm total}
\eeq
The partial rates are
\bea
\label{total_rate_definition}
\Gamma_{ab \to cd}^{\rm total} &=& 
	\int \frac{d^3 p_1 d^3 p_2 d^3 p_3 d^3 p_4 d^3 k_1 d^3 k_2}
 	{64 (2\pi)^{14} \om_1 \om_2 \om_3 \om_4 \Omega_1 \Omega_2}
	\delta^4 (p_1 \!+\! p_2 \!-\! p_3 \!-\! p_4 \!-\! k_1 \!-\! k_2) 
	W_{ab\to cd}(p_1p_2\to p_3p_4k_1k_2) \notag \\
	&~~& \times
\left[ f_a(\om_1) f_b(\om_2) \left(1\!-\!f_c(\om_3)\right) \left(1\!-\!f_d(\om_4)\right) 	\!-\!
	 f_c(\om_1) f_d(\om_2) \left(1\!-\!f_a(\om_3)\right) \left(1\!-\!f_b(\om_4)\right) \right]
\eea
where $a$,$b$,$c$,$d$ are either $e$ or $\mu$,
$W_{ab\to cd}$ is the spin-summed and averaged matrix element.
The charged lepton of flavor $j$ has energy $\om_j = \sqrt{\vp^2_j + m^2_j}$, 
the neutrino of flavor $j$ has energy $\Omega_j = |k_j|$,
and $f_b(\om_j)$ is the Fermi distribution function
\beq
f_b(\om_j) = \left[ 1 + \exp\left( \frac{\om_j - \mu_b}{T} \right)\right]^{-1}
\eeq
Using the previous definitions for $\mu_l$ and $\mu_a$, we have
\beq
\mu_e = \mu_l \!+\! \mu_a, ~~\mu_\mu = \mu_l \!-\! \mu_a
\eeq
and since $\mu_a$ is small, to first order in $\mu_a$ we have
\beq
f_e(\om_1) f_e(\om_2) \left(1\!-\!f_e(\om_3)\right) \left(1\!-\!f_\mu(\om_4)\right) \!-\!
f_e(\om_1) f_\mu(\om_2) \left(1\!-\!f_e(\om_3)\right) \left(1\!-\!f_e(\om_4)\right) 
= F (\om_1, \om_2, \om_3, \om_4) \frac{\mu_a}{T}
\eeq
and
\beq
f_\mu(\om_1) f_e(\om_2) \left(1\!-\!f_\mu(\om_3)\right) \left(1\!-\!f_\mu(\om_4)\right) \!-\!
f_\mu(\om_1) f_\mu(\om_2) \left(1\!-\!f_\mu(\om_3)\right) \left(1\!-\!f_e(\om_4)\right) 
= F (\om_1, \om_2, \om_3, \om_4) \frac{\mu_a}{T}
\eeq
\beq
F (\om_1, \om_2, \om_3, \om_4) \equiv 
  \frac{2 \exp \left[\left(\om_3 \!+\! \om_4 \!-\! 2\mu_l\right)/T \right]
  \left[ 1 \!+\! 2\exp\left[\left(\om_2\!-\!\mu_l\right)/T\right] \!+\!
    \exp\left[\left(\om_2\!+\!\om_4\!-\!2\mu_l\right)/T\right] \right]}
  {\left(1\!+\!\exp[(\om_1\!-\!\mu_l)/T]\right) \left(1\!+\!\exp[(\om_2\!-\!\mu_l)/T]\right)^2
   \left(1\!+\!\exp[(\om_3\!-\!\mu_l)/T]\right) \left(1\!+\!\exp[(\om_4\!-\!\mu_l)/T]\right)^2}
\eeq

\begin{figure}
 \includegraphics[width=0.4\textwidth]{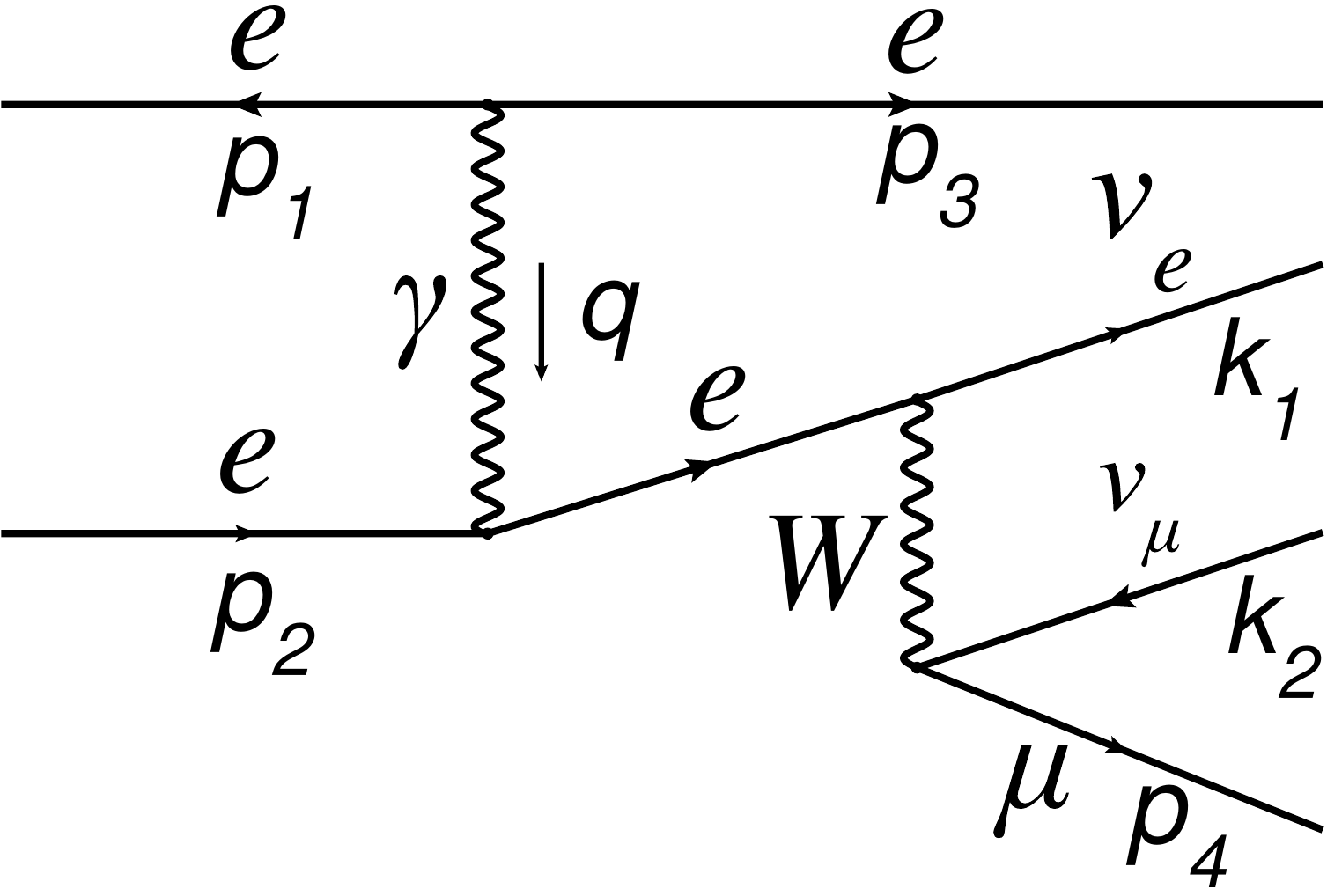}
 \includegraphics[width=0.4\textwidth]{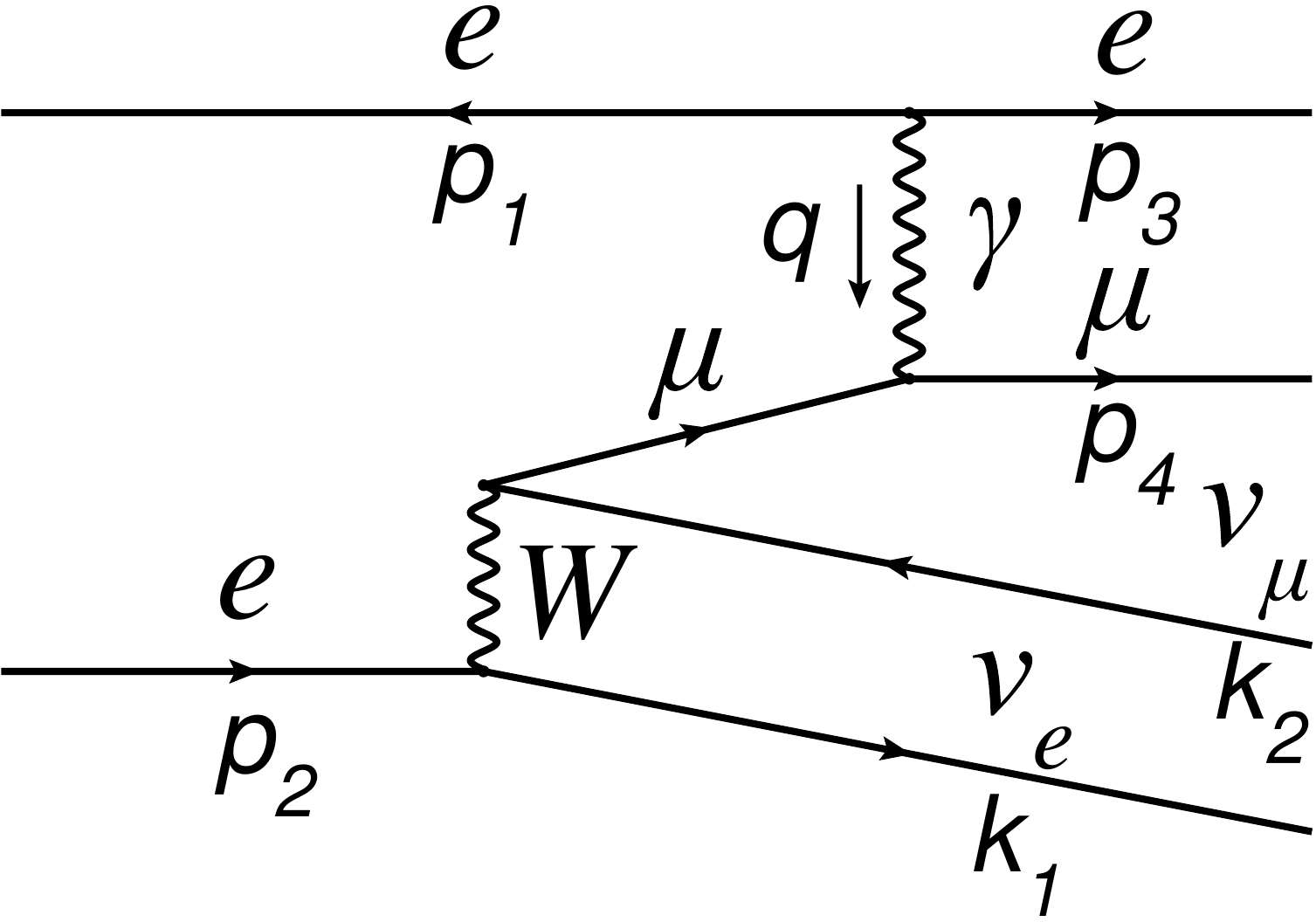}
\caption{Feynman diagrams for the process $e + e \to e + \mu + \numubar + \nue$. There are an additional
         two diagrams which are obtained from these by 
exchanging $p­_1 \leftrightarrow p_2$.}
\label{fig:e_e_to_e_mu}
\end{figure}

\begin{figure}
 \includegraphics[width=0.4\textwidth]{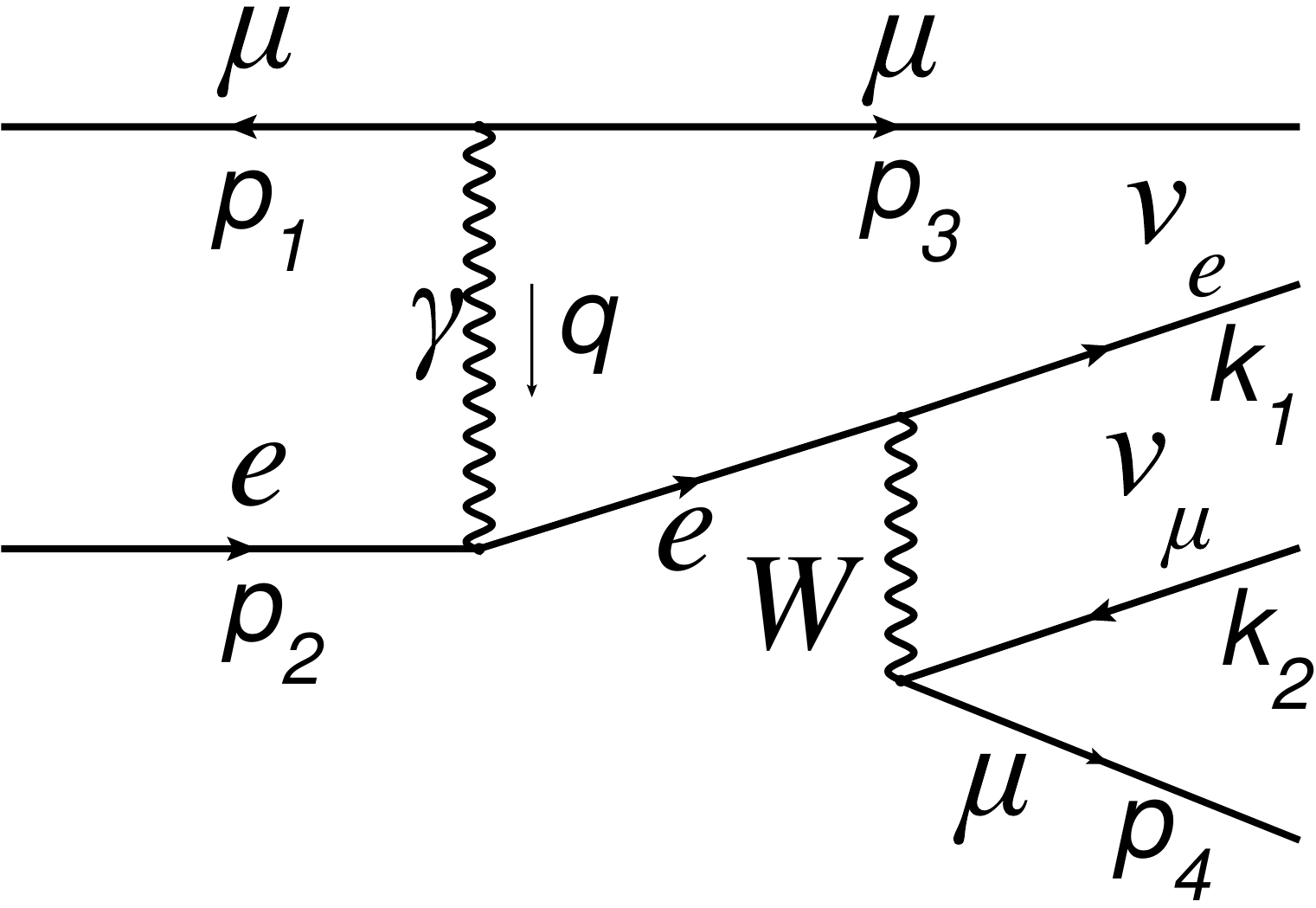}
 \includegraphics[width=0.4\textwidth]{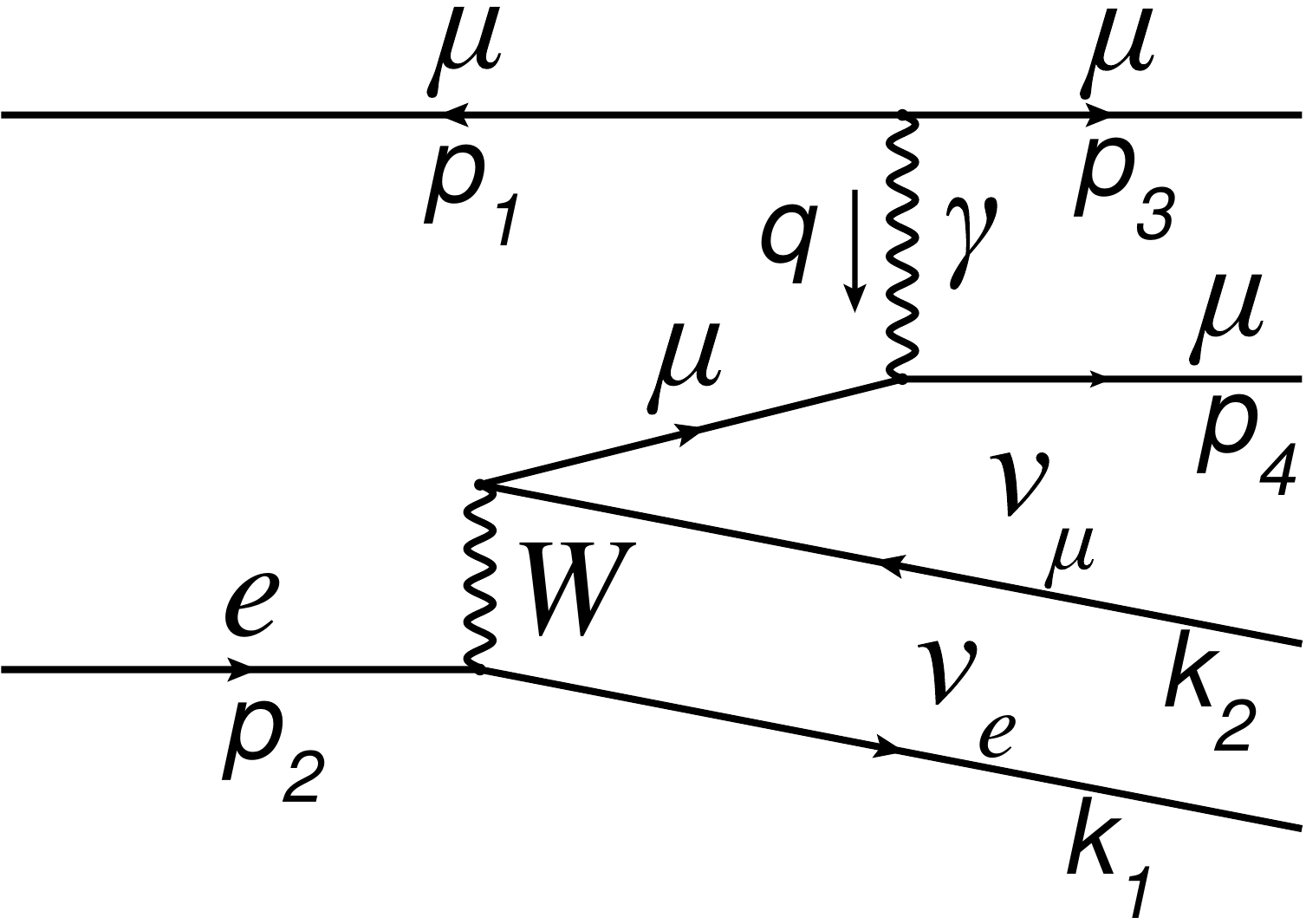}
\caption{Feynman diagrams for the process $e + \mu\to \mu + \mu + \numubar + \nue$. There are an additional
         two diagrams which are obtained from these by 
exchanging $p_3 \leftrightarrow p_4$.}
\label{fig:e_mu_to_mu_mu}
\end{figure}

To determine the content of the matrix elements, we draw the Feynman diagrams for each possible way the
reaction can occur. We can draw two different diagrams for each process, depending on the whether the weak
conversion of the electron to muon occurs before the electromagnetic scattering, or in the reverse order
(Fig. \ref{fig:e_e_to_e_mu}, Fig. \ref{fig:e_mu_to_mu_mu}). However, because there are identical particles involved, and we are integrating over all initial and final momenta, we need to add two additional diagrams for each process. For the process 
$e + e \rightleftharpoons \mu + e + \nu + \bar{\nu}$, we must add two diagrams where the labels on the initial
state electron momenta are reversed; and for the process $e + \mu \rightleftharpoons \mu + \mu + \nu + \bar{\nu}$, we
must add two diagrams where the labels on the final state muon momenta are reversed. These diagrams get an additional
negative sign for the interchange of fermions \cite{Srednicki:2007qs}.
For similar calculations, see \cite{Jaikumar:2005gm,Kaminker:1999sd}.

Since we have four diagrams for each process, the spin summed-and-averaged matrix elements are
\bea
\label{matrix_element_one}
W_{ee\to e\mu} &=& \frac{1}{8} \sum_{spins} |E_1 + E_2 - E_3 - E_4|^2 \notag \\
W_{e\mu \to \mu\mu} &=& \frac{1}{8} \sum_{spins} |M_1 + M_2 - M_3 - M_4|^2
\eea
Here $E_1, E_2, E_3, E_4$ are the amplitudes corresponding to the diagrams of Fig. \ref{fig:e_e_to_e_mu},
and $M_1, M_2, M_3, M_4$ are the amplitudes corresponding to the diagrams of Fig. \ref{fig:e_mu_to_mu_mu} 
\cite{BargerPhillips}:
\bea
E_1 &=& \frac{e^2 G_F}{\sqrt{2}(q^2 \!-\! q^2_s)} \ebar (p_3) \gamma^\mu e(p_1) \nuebar (k_1) \gamma^\lambda
	\projl \frac{\feyn{p}_2 \!+\! \feyn{q} \!+\! m_e}{(p_2\!+\!q)^2 \!-\! m^2_e} \gamma_\mu e(p_2)
	\mubar (p_4) \gamma_\lambda \projl \numu (k_2) \notag \\
E_2 &=& \frac{e^2 G_F}{\sqrt{2}(q^2 \!-\! q^2_s)} \ebar (p_3) \gamma^\mu e(p_1) \nuebar (k_1) \gamma^\lambda
	\projl e(p_2) \mubar (p_4) \gamma_\mu 
	\frac{\feyn{p}_4 \!-\! \feyn{q} \!+\! m_\mu}{(p_4\!-\!q)^2 \!-\! m^2_\mu} 
	\gamma_\lambda \projl \numu (k_2) \notag \\
E_3 &=& \frac{e^2 G_F}{\sqrt{2}(w^2 \!-\! q^2_s)} \ebar (p_3) \gamma^\mu e(p_2) \nuebar (k_1) \gamma^\lambda
	\projl \frac{\feyn{p}_1 \!+\! \feyn{w} \!+\! m_e}{(p_1\!+\!w)^2\!-\! m^2_e} \gamma_\mu e(p_1)
	\mubar (p_4) \gamma_\lambda \projl \numu (k_2) \notag \\
E_4 &=& \frac{e^2 G_F}{\sqrt{2}(w^2 \!-\! q^2_s)} \ebar (p_3) \gamma^\mu e(p_2) \nuebar (k_1) \gamma^\lambda
	\projl e(p_1) \mubar (p_4) \gamma_\mu \frac{\feyn{p}_4 \!-\!\feyn{w}\!+\! m_\mu}{(p_4\!-\!w)^2\!-\!m^2_\mu}
	\gamma_\lambda \projl \numu (k_2) 
\eea

\bea
M_1 &=& \frac{e^2 G_F}{\sqrt{2}(q^2 \!-\! q^2_s)} \mubar (p_3) \gamma^\mu \mu(p_1) \nuebar (k_1) \gamma^\lambda
	\projl \frac{\feyn{p}_2 \!+\! \feyn{q} \!+\! m_e}{(p_2\!+\!q)^2 \!-\! m^2_e} \gamma_\mu e(p_2)
	\mubar (p_4) \gamma_\lambda \projl \numu (k_2) \notag \\
M_2 &=& \frac{e^2 G_F}{\sqrt{2}(q^2 \!-\! q^2_s)} \mubar (p_3) \gamma^\mu \mu(p_1) \nuebar (k_1) \gamma^\lambda
	\projl e(p_2) \mubar (p_4) \gamma_\mu 
	\frac{\feyn{p}_4 \!-\! \feyn{q} \!+\! m_\mu}{(p_4\!-\!q)^2 \!-\! m^2_\mu} 
	\gamma_\lambda \projl \numu (k_2) \notag \\
M_3 &=& \frac{e^2 G_F}{\sqrt{2}(s^2 \!-\! q^2_s)} \mubar (p_4) \gamma^\mu \mu(p_1) \nuebar (k_1) \gamma^\lambda
	\projl \frac{\feyn{p}_2 \!+\! \feyn{s} \!+\! m_e}{(p_2\!+\!s)^2\!-\! m^2_e} \gamma_\mu e(p_2)
	\mubar (p_3) \gamma_\lambda \projl \numu (k_2) \notag \\
M_4 &=& \frac{e^2 G_F}{\sqrt{2}(s^2 \!-\! q^2_s)} \mubar (p_4) \gamma^\mu \mu(p_1) \nuebar (k_1) \gamma^\lambda
	\projl e(p_2) \mubar (p_3) \gamma_\mu \frac{\feyn{p}_3 \!-\!\feyn{s}\!+\! m_\mu}{(p_3\!-\!s)^2\!-\!m^2_\mu}
	\gamma_\lambda \projl \numu (k_2)
\eea
where $w = p_2 - p_3$, and $s = p_1 - p_4$.

The only parameter in our calculation that depends on details of the
baryonic matter in the neutron star is the plasma screening momentum $q_s$.
In a full treatment one would have to use the appropriate
in-medium propagator which is a complicated function of the photon momentum.

In this paper we greatly simplify the calculation by
assuming that the
longitudinal and transverse photons have a common screening mass 
\beq
q_s^2 = 5 \al \mu_l^2 \ .
\label{qs}
\eeq
We argue in Appendix~\ref{sec:screening} that this leads to
an estimate of the bulk viscosity that is correct 
to within an order of magnitude at reasonable densities for nuclear matter.
As a further test we also performed calculations with no 
screening at all ($q_s^2=0$) and found that the bulk viscosity 
shifted by no more than one order of magnitude.

To obtain the equilibration rates,
we first multiply out the right hand sides of \eqn{matrix_element_one} and define partial matrix elements by
\bea
W_{ee\to e\mu} &=& \sum_{i,j\leq i} W^{ij}_{ee\to e\mu},~ 
  W_{e\mu\to \mu\mu} = \sum_{i,j\leq i} W^{ij}_{e\mu\to \mu\mu} \notag \\
  W^{11}_{ee\to e\mu} &=& \frac{1}{8} \sum_{spins} |E_1|^2,~
  W^{12}_{ee\to e\mu} = \frac{1}{8} \sum_{spins} (E^\dag_1 E_2 + E^\dag_2 E_1),~ 
  W^{13}_{ee\to e\mu} = -\frac{1}{8} \sum_{spins} (E^\dag_1 E_3 + E^\dag_3 E_1),~ {\rm etc.} 
\eea
The traces resulting from the spin sums are easily evaluated with a computer algebra package; we used the FeynCalc package for Mathematica \cite{feyncalc}. In the next few paragraphs,
we will describe the steps used to analytically integrate 10 of the 18 integrals, and list the expressions that we subsequently integrated numerically in 
Appendix \ref{sec:pri}.

We make use of the fact that the neutrino energies are $\sim T \ll
\mu_e, \mu_\mu$ by approximating the momentum and energy conserving
delta functions as
\beq
\delta^4 (p_1\!+\!p_2\!-\!p_3\!-\!p_4\!-\!k_1\!-\!k_2) \approx 
	\delta(\om_1\!+\!\om_2\!-\!\om_3\!-\!\om_4\!-\!\Omega_1\!-\!\Omega_2) \delta^3 (\vp_1\!+\!\vp_2\!-\!\vp_3\!-\!\vp_4) \ .
\label{delta}
\eeq 
We then note that $k_1$ and $k_2$ occur exactly once in each term, dotted into one of the other 4-momenta $p_i$. Writing
\beq
k_j = \Omega_j \bigl( 1, \sin\xi_j \cos\eta_j, \sin\xi_j \sin\eta_j, \cos\xi_j \bigr)
\eeq
we can see that any dot product with another 4-momentum $p_i$ is 
\beq
p_i \cdot k_j = \Omega_j \bigl( \om_i \!-\! (p_i)_x \sin\xi_j \cos\eta_j \!-\! (p_i)_y \sin\xi_j \sin\eta_j
	\!-\! (p_i)_z \cos\xi_j \bigr) \ .
\eeq
The integrals over the $k_1$ and $k_2$ angular variables then become trivial:
\beq
\int \frac{d^3 k_j}{\Omega_j} p_i \cdot k_j =
	\int k^2_j dk_j d(\cos \xi_j) d\eta_j p_i \cdot \hat{k}_j
	= 4\pi \om_i \int_0^\infty \Omega^2_j d\Omega_j
\eeq
because all of the integrations over one of the angles $\xi_j$ or $\eta_j$ are zero.

The energy-momentum conserving delta function allows us to use relations like $p_1-p_3 = p_4 - p_2$ to rewrite
some of the denominators of the matrix elements. For example, in $W_{ee\to e\mu}$ we can substitute variables so that
$p_3$ does not appear in the denominators of any of the terms; then we can integrate out the $p_3$ 3-momentum variables easily. Similarly, in $W_{e\mu\to \mu\mu}$ we can substitute variables so that $p_2$ does not appear in the denominators and
integrate out the $p_2$ 3-momentum variables. However, our matrix elements have many terms containing the 
four-momentum $p_3$ ($p_2$), so it would be easier if we could integrate over $d^4 p_3$ ($d^4 p_2$). This is accomplished by replacing
\beq
\int \frac{d^3 p_3}{\om_3} = 
	\int \frac{d^4 p_3}{(p_3)_0} \delta \left((p_3)_0 - \sqrt{\vp^2_3 + m^2_\mu} \right) 
\approx \int \frac{d^4 p_3}{(p_3)_0} \delta \left((p_3)_0 - \mu_l\right) 
\eeq
in $W_{ee\to e\mu}$ and similarly for $p_2$ in $W_{e\mu\to \mu\mu}$ . In the last approximation we are using the fact that the Fermi distribution function is sharply peaked at low temperatures. Then we integrate over $d^4 p_3$ ($d^4 p_2$) using four of the delta functions. 

We can further approximate that the medium is isotropic, by taking one of the remaining momentum variables to be in a fixed direction (the $z$-axis for convenience). The electrons are relativistic, so $\om_i = |\vp_i|$ and 
$d^3 p_i = \om^2_i d\om_i d\cos \theta_i d\phi_i$ when particle $i$ is an electron. The muons may not be relativistic,
so $\om_i = \sqrt{\vp_i^2 + m_\mu^2}$ and $d^3 p_i = \om_i \sqrt{\om_i^2\!-\!m_\mu^2} d\om_i d\cos\theta_i d\phi_i$ when
particle $i$ is a muon. We then use the remaining delta function to integrate over the magnitude of this isotropic momentum
variable.

The remainder of the integrations are performed numerically. The only further approximation made was to 
again take advantage of the sharply peaked Fermi distribution function, and set $\om_i = \mu_l$ everywhere inside
the integral, except for inside the Fermi function itself. This allows a separation of the eight-dimensional integral
into a four-dimensional energy integral and a four-dimensional integral over the angular variables. The integration
variables are also changed to dimensionless variables by scaling them with respect to $\mu_l$. 

The final expression for each term in the rate has the form
\beq
\Gamma_{e \ell\to \mu \ell}^{ij} = 
\frac{e^4 G^2_F \mu_l^{12}}{128 \pi^{11} m_\mu^4} \left(\frac{\mu_a}{T}\right) \times I^\ell_\om \times I^{\ell i j}_{d\Omega}
\eeq
where $\ell$ is the species of the spectator lepton, and
$I^\ell_\om$ and $I^{\ell i j}_{d\Omega}$ are dimensionless energy and angular 
integrals, respectively. These integrals are listed in Appendix \ref{sec:pri}.

\section{Numerical results and conclusions}
\label{sec:results}

The remaining part of the rate calculation is performed numerically. The dimensionless energy integrals are nearly the same; a power-law fit of the results yields
\beq
I^e_\om \approx 78.86 \left( \frac{T}{\mu_l} \right)^8, ~~~~
I^\mu_\om \approx 78.62 \left( \frac{T}{\mu_l} \right)^8
\eeq

In our approximation, the angular integrals only have dependence on $\mu_l$. 
We determined an analytical fit for
the $\mu_l$-dependence of $I^{e ij}_{d\Omega}$ and $I^{\mu ij}_{d\Omega}$ (accurate within 5\%) over the range $120\,\MeV < \mu_l < 300\,\MeV$ 
by curve-fitting the numerical data with sixth-order polynomials:
\bea
\sum_{ij} I^{e ij}_{d\Omega} &\approx& \left(1-\frac{m^2_\mu}{\mu^2_l}\right)^{1/2}
\sum_{i = 0}^6 c_i \left(\frac{\mu_l}{m_\mu}\right)^i, \notag \\
c_0 &=& -1.7363 \times 10^4, c_1 = 5.0189 \times 10^4, c_2 = -4.7644 \times 10^4,
   c_3 = 1.3224 \times 10^4, \notag \\
c_4 &=& 4.4203 \times 10^3, c_5 = -2.7199 \times 10^3, c_6 = 3.5119 \times 10^2
\eea
\bea
\sum_{ij} I^{\mu ij}_{d\Omega} &\approx& \left(1-\frac{m^2_\mu}{\mu^2_l}\right)^{3/2}
\sum_{i = 0}^6 c_i \left(\frac{\mu_l}{m_\mu}\right)^i, \notag \\
c_0 &=& 1.2433 \times 10^6, c_1 = -3.6329 \times 10^6, c_2 = 4.4365 \times 10^6,
   c_3 = -2.8702 \times 10^6, \notag \\
c_4 &=& 1.0354 \times 10^6, c_5 = -1.9728 \times 10^5, c_6 = 1.5507 \times 10^4
\eea

\begin{figure}
 \includegraphics[width=0.8\textwidth]{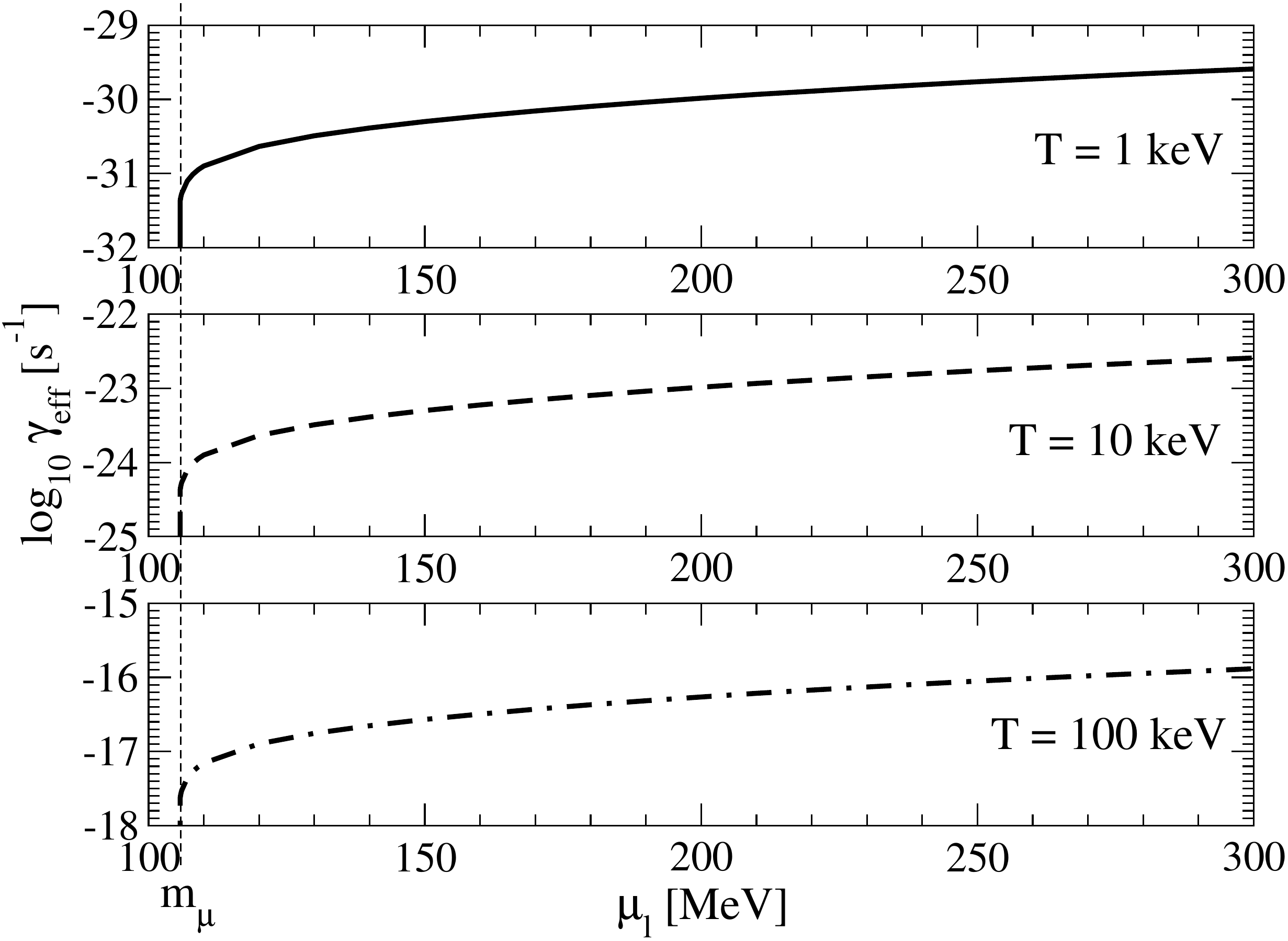}
\caption{
Dependence of the effective rate of electron/muon conversion
$\gaeff$ (see \eqn{C_gamma}) on the charged-lepton chemical potential
$\mu_l$ at three different
temperatures.
As $\mu_l$ drops towards $m_\mu$, the muon population decreases and
the conversion rate drops to zero. The temperature dependence
is $T^7$, hence $\gaeff$ is much larger at higher temperatures.
}
\label{fig:gammaeff_vs_mu}
\end{figure}

Fig.~\ref{fig:gammaeff_vs_mu} shows the $\mu_l$ dependence of the
effective rate $\gaeff$ defined in \eqn{C_gamma}. As $\mu_l$
approaches $m_\mu$, the rate quickly drops to zero as the muon population
disappears.  The overall $T^7$ dependence is also
illustrated in the sizable difference in order of magnitude of the
rate for the three different temperatures.

\begin{figure}
 \includegraphics[width=0.8\textwidth]{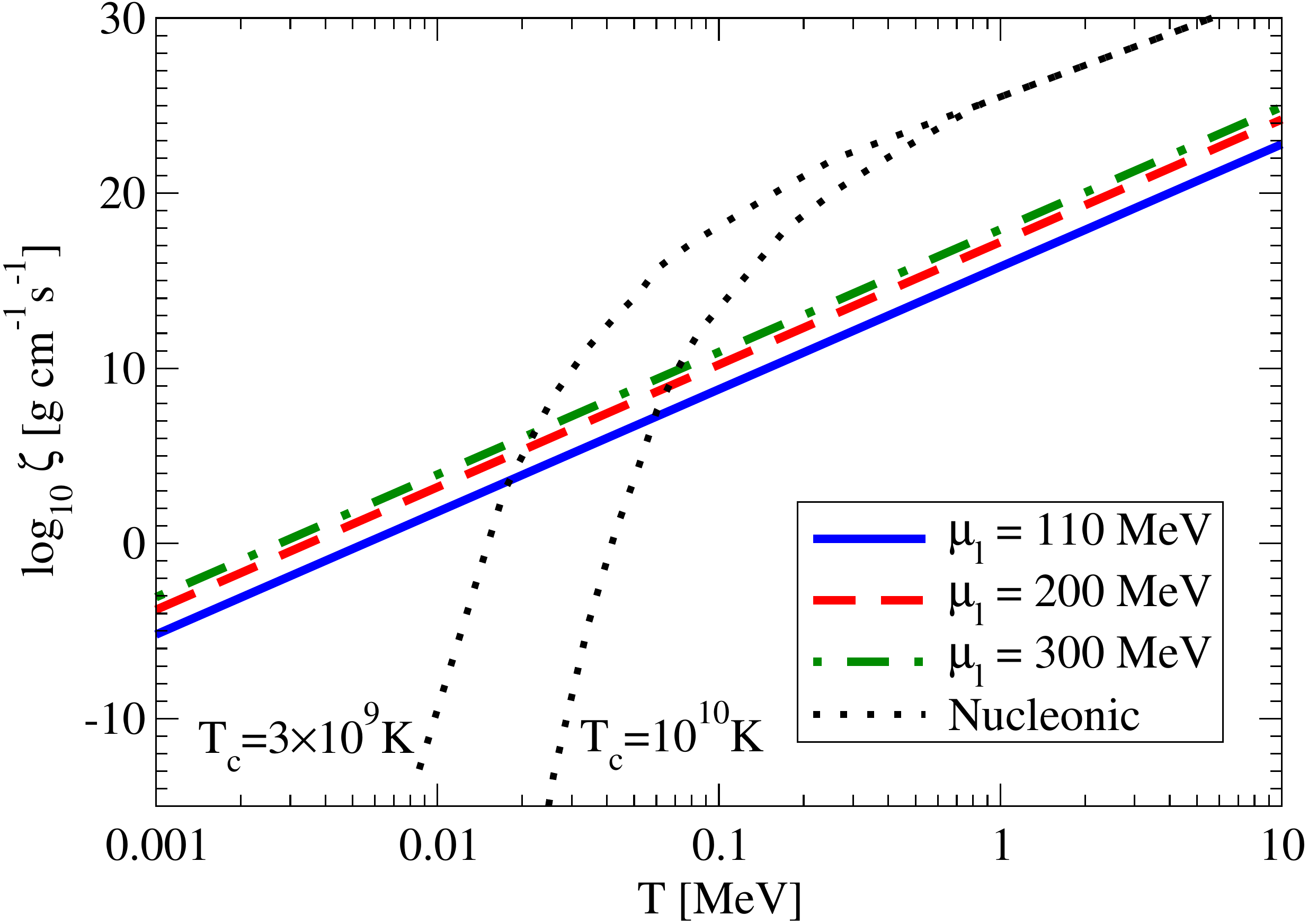}
\caption{(Color online) 
Dependence of the leptonic bulk viscosity $\zeta$ on temperature
for three different
values of the lepton chemical potential,
and an oscillation frequency of 1\,{\rm kHz};
for frequency dependence, see the discussion after \eqn{zeta-slow}.
We also show
the nucleonic bulk viscosity \cite{Haensel:2001mw-ADS} due to
modified-Urca processes, for two values of the critical temperature.
}
\label{fig:bulkviscosity_vs_T}
\end{figure}

Fig.~\ref{fig:bulkviscosity_vs_T} shows the temperature dependence of
the leptonic bulk viscosity $\zeta$ as defined in \eqn{zeta_phi}, for an
oscillation frequency $\om = 2\pi \times 1 {\rm kHz}$. The
three approximately straight lines on the log-log plot
illustrate the power-law dependence on $T$ for three different values of
$\mu_l$. 
Also plotted are dotted curves showing the nucleonic bulk viscosity 
for two different values of the critical temperature $T_c$. These are
obtained from Ref.~\cite{Haensel:2001mw-ADS} in a model
where the neutrons are superfluid, pairing in the spin triplet state,
the protons are superconducting,
pairing in the spin singlet state, and they have
a common critical temperature $T_c=T_{cp}=T_{cn}$.
Also, it is assumed that only modified Urca processes
are available for damping of pulsations (although direct Urca processes 
would become possible at higher densities).
Above the critical temperature for superfluidity/superfluidity,
the bulk viscosity for 1~kHz oscillations
due to leptons is several orders
of magnitude less than the bulk viscosity due to nucleons. Below the
critical temperature, the nucleonic bulk viscosity quickly decreases
and at a low enough temperature, the leptonic contribution becomes
dominant. Based on our calculations, this crossover temperature
appears to be of order 0.01 to 0.1 MeV ($10^8$ to $10^9$~K)
for an oscillation frequency in the kHz
range. Such a suppression of the nucleonic contribution can arise
either from superfluidity of neutrons or from superconductivity
of protons. It is therefore quite possible that for many cold
neutron stars, the
bulk viscosity of the superconducting or superfluid region comes mainly
from leptonic processes. 
In regions that are neither superconducting nor superfluid
(more strictly, where $T \gtrsim T_{cp}$ and $T \gtrsim T_{cn}$)
the nucleonic bulk viscosity will likely dominate.

The viscosity curves in Fig.~\ref{fig:bulkviscosity_vs_T}
all slope upwards because the equilibration rate $\gaeff(T)$
is well below the oscillation frequency $\om$, so we are in the
slow-equilibration (high frequency) regime of \eqn{zeta_phi},
where 
\beq
\zeta \approx C \frac{\gaeff(T)}{\om^2} \ .
\label{zeta-slow}
\eeq
This is true for both leptonic and nuclear viscosities.
In this regime one can simply add the two bulk viscosities to get the total
bulk viscosity (see, for example, appendix A of Ref.~\cite{Alford:2006gy}).
As the temperature rises, the equilibration rate and hence the
bulk viscosity rise. When $\gaeff(T)$ comes close to $\om$, \eqn{zeta-slow}
becomes a poor approximation to \eqn{zeta_phi}: 
$\zeta$ reaches a maximum when $\gaeff(T)=\om$.
Those maxima, for both leptonic and nuclear bulk viscosities,
are beyond the right hand limit of Fig.~\ref{fig:bulkviscosity_vs_T};
for $\mu_l=200~\MeV$, the peak occurs at $T\approx 40~\MeV$.

We can now see how our results depend on the frequency of the
oscillations.  Decreasing $\om$ moves each $\zeta(T)$ curve to the
left, shifting the viscosity curves in
Fig.~\ref{fig:bulkviscosity_vs_T} upwards.  
The largest value we find for the leptonic effective rate 
(at $T=10~\MeV$, for $\mu_l=300~\MeV$) is $\gaeff\sim 2$~rad/s,
so for the leptonic bulk viscosity
\eqn{zeta-slow} is valid for oscillation frequencies well above this
value.  For example, if we reduced the oscillation frequency from
1000~Hz to 100~Hz then all the viscosity curves in
Fig.~\ref{fig:bulkviscosity_vs_T} would be shifted upwards by a factor
of $100$.  Decreasing the frequency still further would bring us
to the regime where, in the temperature range of interest, either
the nuclear and leptonic rate was comparable to the oscillation
frequency (so one or both bulk viscosity curves would show a
resonant peak in our plot).  Then one may not be able to simply add
the bulk viscosities.  At extremely low oscillation frequencies,
both peaks would shift to very low temperatures, the bulk viscosity
curves in our plot would all slope downwards, the nucleonic
contribution dominates, and the bulk
viscosities could again be added.

It will be interesting to see whether the leptonic contribution that we have
calculated here has any impact on oscillations of neutron stars. In the case
of r-modes, shear viscosity becomes the dominant source of damping in the low
temperature regime, so the leptonic contributions to the bulk viscosity at low
temperature are not likely to be an important source of r-mode damping.
Also the shear viscosity $\eta$ of superfluid nuclear matter is much
larger than the leptonic bulk viscosity we have calculated: $\eta\sim
10^{16}\,{\rm g}\,{\rm cm}^{-1}{\rm s}^{-1}$ at $T\sim 0.1\,\MeV$, rising
to $\eta\sim 10^{22}\,{\rm g}\,{\rm cm}^{-1}{\rm s}^{-1}$ at $T\sim
0.001\,\MeV$ (see Fig.~5 of Ref.~\cite{Shternin:2008es}) so bulk viscosity
would only dominate the damping of modes with very little shear flow.
Radial pulsations \cite{Cutler:1990,Gusakov:2006ga} would be an interesting
example to investigate.
We used a rough approximation \eqn{qs} to treat the photon screening;
we argued (Appendix~\ref{sec:screening})
that this is valid to within about an order of magnitude, but
if a more precise estimate of the
bulk viscosity were required, one could improve on our treatment
by replacing the approximation \eqn{qs} with separate propagators
for the transverse and longitudinal photons, incorporating their
separate screening mechanisms \cite{Shternin:2007ee}.
It should be noted that our calculation
is limited to the small-amplitude regime ($\mu_a\ll T$). If the
leptonic bulk viscosity is insufficient to damp an unstable oscillation
such as an $r$-mode then the amplitude will rise and it will be
necessary to repeat our calculation in the large-amplitude (``supra-thermal'')
regime \cite{Haensel:2002qw}
to see whether leptonic bulk viscosity can stop the growth
of the mode once it reaches a large enough amplitude.

\section*{Acknowledgements}
We thank Sanjay Reddy for initial discussions that led to this project.
We thank Peter Shternin, Dima Yakovlev, and the anonymous referee
for many helpful comments.
This research was
supported in part by the Office of Nuclear Physics 
and High Energy Physics 
of the U.S.~Department of Energy under contracts
\#DE-FG02-91ER40628,  
\#DE-FG02-05ER41375. 

\appendix

\section{Photon screening}
\label{sec:screening}

In this appendix we discuss the adequacy of our approximation \eqn{qs} for the
internal photon propagator in the modified Urca process for leptons.
The energy $\om$ of the photon is $\sim T$ because
all the initial and final state particles have energies within $T$ of
their Fermi energies; however, the photon 3-momentum $q$ must be
large enough to move a lepton between the muon and electron Fermi surfaces,
so $q\geqslant \qmin$ where $\qmin=p_{F,e}-p_{F,\mu}$. Thus $\om\ll q$
and we can write the photon self-energy in the static limit where it
only depends on $q$.
There are contributions to the longitudinal and transverse self-energies from 
protons, electrons, and muons. If the protons are
superconducting, as they are at the temperatures of interest in this
paper, then they provide an additional contribution to the transverse
photon self-energy. The complete expressions are
\beq
\ba{rcl}
\Pi_L(q) &=&\dsp M^2_{D,p} \,\xi_{L}\Bigl(\frac{q}{k_{F,p}}\Bigr)
 + M^2_{D,e}\,\xi_{L}\Bigl(\frac{q}{k_{F,e}}\Bigr)
 + M^2_{D,\mu}\,\xi_{L}\Bigl(\frac{q}{k_{F,\mu}}\Bigr) \ , \\[3ex]
\Pi_T(q) &=&\dsp M^2_{D,p} \,\xi_{T}\Bigl(\frac{q}{k_{F,p}}\Bigr)
 + M^2_{D,e}\,\xi_{T}\Bigl(\frac{q}{k_{F,e}}\Bigr)
 + M^2_{D,\mu}\,\xi_{T}\Bigl(\frac{q}{k_{F,\mu}}\Bigr)
 + \Pi^{\rm (sc)}_p(q) \ .
\ea
\label{polarization}
\eeq
The Debye mass for a given species is (see, for example, \cite{Shternin:2006uq})
\beq
M_D^2=4\pi\al \mu k_F
\label{MDebye}
\eeq
where $\mu$ is the Fermi energy (defined relativistically,
so $\mu^2 = k_F^2 + m^2$) and $k_F$ is the Fermi momentum. 
The screening functions $\xi_L$ and $\xi_T$ in the static limit are 
real, and are given by
\newcommand{\qb}{\bar q}
\beq
\ba{rcl}
\xi_L(\qb) &=&\dsp \frac{1}{2}
 + \frac{1}{2\qb}\Bigl(1-\frac{\qb^2}{4}\Bigr)
 \log \Bigl| \frac{\qb+2}{\qb-2} \Bigr| \ , \\[3ex]
\xi_T(\qb) &=&\dsp \frac{1}{8}\biggl(
 1 + \frac{\qb^2}{4} - \frac{1}{\qb}\Bigl(1-\frac{\qb^2}{4}\Bigr)^2
 \log \Bigl| \frac{\qb+2}{\qb-2} \Bigr| \biggr) \ .
\ea
\label{xi}
\eeq
The full expressions for photon screening by a degenerate gas of charged
fermions were first obtained by Lindhard \cite{Lindhard:1954}.
Eq.~\eqn{xi} was obtained from the version of
Lindhard's expressions for the
dielectric permittivities $\ep_L$ and $\ep_T$ 
given in Ref.~\cite{Cockayne:2006},
using the fact that $\Pi_L(\om,q) = (\om^2-q^2)(1 - \ep_L)$ and
$\Pi_T(\om,q) = \om^2(1 - \ep_T)$ (see Sec. (6.4) of Ref.~\cite{LeBellac}).
Note that $\xi_L$ and $\xi_T$ above are defined
in the static limit, where $\om/q\to 0$ at fixed $q$. They are therefore 
different from the quantities $\chi_l(x)$ and $\chi_t(x)$ which are
commonly given in the literature \cite{Shternin:2007ee,Heiselberg:1993cr},
and are calculated at $\om=xq$ in the limit $q\to 0$.

In our calculations of the leptonic flavor equilibration rate
we use the rough approximation $\Pi_L=\Pi_T=q_s^2$ 
\eqn{qs} instead of the correct screening expressions given above.
We now explain why this is a reasonable approximation.

First we discuss the longitudinal photons. Their momentum varies from
$\qmin$ up to $k_{F,e}+k_{F,\mu}$, but the momentum dependence
of $\Pi_T$ is very moderate: from \eqn{xi} we see that as $\qb$ varies
from 0 to 2, $\xi_L$ varies from 1 to $\half$. In order to judge whether,
for the denominator of the longitudinal photon propagator,
$q^2 + q_s^2$ is a good approximation to $q^2 + \Pi_L(q)$,
we use a naive free particle model for nuclear matter. 
In Table~\ref{tab:screening} we show the results.
At each value of the baryon chemical potential $\mu_B$, the
negative-charge chemical potential $\mu_l$ is determined by requiring
overall electrical neutrality. This then fixes the Fermi momenta of
the protons, electrons, and muons. In Table~\ref{tab:screening}
we see that when $q=\qmin$ (which is where there is greatest sensitivity
to the exact form of the screening), the difference between 
$q^2 + q_s^2$ and $q^2 + \Pi_L(\qmin)$ is a few percent at low density,
and still less than a factor of 2 at very high densities.

For the transverse photons, $\xi_T$ varies from 0 at $q=0$ to
$\frac{1}{3}$ at $q=\infty$, so the normal-fermion contribution
to the transverse screening is more important at higher momenta.
The other contribution to $\Pi_T$ comes from the superconducting protons,
and it is more important at low momentum.
At zero momentum we have Meissner screening, but as the momentum
rises the effective screening mass drops slowly: this is
seen in the calculation of Ref.~\cite{Shternin:2007ee}
which finds that, for $q\gg \xi^{-1}$ (where 
the correlation length $\xi=p_{F,p}/(m_pT_{c,p})$), and assuming the
static limit,
\beq
\Pi^{(sc)}_T(q) \approx \frac{\pi \al p_{F,p}^2 T_{cp}}{ q} \ .
\label{PiT_sc}
\eeq
(This result follows from Ref.~\cite{Shternin:2007ee} eqn (49),
taking $\om\to 0$ and using $Q=\pi^2$ as specified in the
preceding paragraph.)
In Table~\ref{tab:screening} we show numerical results for the 
naive free-nucleon
model of nuclear matter. We assumed $T_{cp}=1~\MeV$ 
(see Ref.~\cite{Muther:2005cj}, fig.~10,
and  Ref.~\cite{Baldo:2007jx}, fig.~2).
At the lowest allowed photon momentum $q=\qmin$, 
which is where there is greatest sensitivity to the exact form of the 
screening, the difference between 
$q^2 + q_s^2$ and $q^2 + \Pi_T^2$ is a few percent at low density, but
rises to a factor of 3 at density $n/n_{\rm sat}=12$, and a factor
of 10 at $n/n_{\rm sat}=27$. 

We conclude that our rough approximation of using a
photon self-energy $q_s^2=5\al\mu_l^2$ \eqn{qs} gives
a reasonable estimate of the in-medium photon propagator.
At low densities it is accurate to within 10\%.
At higher densities, up to 10 times nuclear saturation density
in the simple model of Table~\ref{tab:screening}, 
our approximation underestimates the
screening of longitudinal photons by a factor of about 2
and overestimates the screening of transverse photons by a
factor of about 3. (At even higher densities, where a description
in terms of nucleons is probably no longer appropriate, 
our approximation for transverse screening deviates further from 
the free nucleon model.)
Since the rate involves the square of the photon propagator, we
conclude that our approximate treatment of the photon propagator affects the
rate by less than an order of magnitude at reasonable densities for nuclear
matter.

\begin{table}[htb]
\def\st{\rule[-1ex]{0em}{3ex}} 
\begin{tabular}{c@{\quad}c@{\quad}c@{\quad}c@{\quad}c@{\quad}c@{\quad}c@{\quad}c@{\quad}c@{\quad}c}
\st
$\mu_B$ & $n/n_{\rm sat}$ & $\mu_l$ & $\qmin^2$ & $\Pi_L(\qmin)$ & $\Pi_T(\qmin)$ & $q_s^2=5\al\mu_l^2$ 
 &  $\qmin^2+\Pi_L(\qmin)$ &  $\qmin^2+\Pi_T(\qmin)$ & $\qmin^2+q_s^2$ \\
\hline
\st 1056 & 3.164 & 111.1 & 5908 & 1067 & 55.45 & 450.1 & 6974 & 5963 & 6358 \\
\st 1125 & 6.76 & 167.6 & 1406 & 2150 & 30.13 & 1025 & 3557 & 1436 & 2431 \\
\st 1200 & 12.03 & 224.4 & 698.3 & 3300 & 65.06 & 1838 & 3999 & 763.4 & 2537 \\ 
\st 1350 & 26.93 & 328.7 & 304.3 & 5783 & 215 & 3943 & 6087 & 519.3 & 4247 \\
\end{tabular}
\caption{Screening parameters in MeV or MeV$^2$
for a free-nucleon model of $npe\mu$ nuclear
matter; $\mu_B$ is the baryon number chemical potential, $n/n_{\rm sat}$
is the baryon density relative to nuclear saturation density; $\mu_l$
is the Fermi energy of the electrons and muons; 
$\qmin$ is the lowest photon momentum that contributes to the modified
Urca process; $\Pi_L$ and $\Pi_T$
are defined in \eqn{polarization}. The last three columns compare our
approximate photon propagator at $q=\qmin$
(final column) with the photon propagator using the full screening
expressions given in Appendix~\ref{sec:screening}.
}
\label{tab:screening}
\end{table}

\section{Partial Rate Integrals}
\label{sec:pri}

The following abbreviations are used throughout this appendix:
\bea
x_m &=& \frac{m_\mu}{\mu_l},~ x_s = \frac{q_s}{\mu_l},~ t = \frac{T}{\mu_l} \notag \\
C_{12} &=& 1-\cos\theta_2,~ C_{14} = 1-\sqrt{1-x_m^2}\cos\theta_4 \notag \\
C_{24} &=& 1-\sqrt{1-x_m^2}\left( \sin\theta_2 \sin\theta_4(\sin\phi_2 \sin\phi_4 \!+\! \cos\phi_2 \cos\phi_4) 
                                 \!+\! \cos\theta_2 \cos\theta_4 \right) \notag \\
\barC_{13} &=& 1-(1-x_m^2)\left( \sin\theta_1 \sin\theta_3(\sin\phi_1 \sin\phi_3 \!+\! \cos\phi_1 \cos\phi_3) 
                                 \!+\! \cos\theta_1 \cos\theta_3 \right) \notag \\
\barC_{14} &=& 1-(1-x_m^2)\cos\theta_1,~ \barC_{34} = 1-(1-x_m^2)\cos\theta_3 \notag \\
F\left(x_a, x_b, x_c, x_d \right) &=& 
  \frac{2 \exp \left[\left(x_c \!+\! x_d \!-\! 2\right)/t \right]
  \left[ 1 \!+\! 2\exp\left[\left(x_b\!-\!1\right)/t\right] \!+\!
    \exp\left[\left(x_b\!+\!x_d\!-\!2\right)/t\right] \right]}
  {\left(1\!+\!\exp[(x_a\!-\!1)/t]\right) \left(1\!+\!\exp[(x_b\!-\!1)/t]\right)^2
   \left(1\!+\!\exp[(x_c\!-\!1)/t]\right) \left(1\!+\!\exp[(x_d\!-\!1)/t]\right)^2}
\eea

\bea
I^e_\om &=& \int dx_2 dx_4 dy_1 dy_2 ~y_1^2 ~y_2^2 ~F\left(x_4\!+\!y_1\!+\!y_2\!-\!x_2\!+\!1, x_2, 1, x_4 \right) \notag \\
I^\mu_\om &=& \int dx_1 dx_3 dy_1 dy_2 ~y_1^2 ~y_2^2 ~F\left(x_1, 1, x_3, x_1\!-\!x_3\!-\!y_1\!-\!y_2\!+\!1 \right)
\eea

\bea
I^{e11}_{d\Omega} &=& \sqrt{1-x­_m^2}
\int d\Omega_2 d\Omega_4 \frac{4C_{12} C_{14} \!+\! 2C_{12} C_{24} \!-\! 2C_{14}C_{24} \!-\! 4x_m^2 C_{12} \!+\! x_m^2 C_{24}}
                              {(x_m^2 \!-\! 2C_{24} \!-\! x_s^2)^2}
\eea
\bea
I^{e12}_{d\Omega} &=& \!-\!\sqrt{1-x­_m^2}
\int d\Omega_2 d\Omega_4 \Biggl[ \frac{\!-\!2C_{12} C_{24}^2 \!+\! 2C_{14} C_{24}^2 \!+\! 8C_{12} C_{14} \!+\! 4C_{12} C_{24}
                               \!-\!4C_{12} C_{14} C_{24} \!-\! 4C_{14}C_{24}}
                              {(x_m^2 \!-\! 2C_{24} \!-\! x_s^2)^2} \notag \\ &~&~~~~~~~~~~~~~~~~~~~~~~~~~+
                                 \frac{x_m^2 \Bigl(4C_{12}^2 \!-\! 8C_{12}
                                       \!+\! 4C_{14} \!+\! C_{12}C_{24} \!-\! C_{14}C_{24}\Bigr) - x_m^4}
                              {(x_m^2 \!-\! 2C_{24} \!-\! x_s^2)^2} \Biggr]
\eea
\bea
I^{e13}_{d\Omega} &=& \!-\!\sqrt{1-x­_m^2}
\int d\Omega_2 d\Omega_4 \frac{\!-\!4C_{12} C_{14} \!-\!4C_{12} C_{24} \!+\! 6x_m^2 C_{12}}
                              {(x_m^2 \!-\! 2C_{24} \!-\! x_s^2)(x_m^2 \!-\! 2C_{14} \!-\! x_s^2)}
\eea
\bea
I^{e14}_{d\Omega} &=&  \sqrt{1-x­_m^2}
\int d\Omega_2 d\Omega_4 \Biggl[\frac{2C_{12} C_{14}^2 \!-\!4C_{12} C_{14} \!-\!4C_{12} C_{24} \!+\! 2C_{12} C_{14} C_{24}}
                                     {(x_m^2 \!-\! 2C_{24} \!-\! x_s^2)(x_m^2 \!-\! 2C_{14} \!-\! x_s^2)} \notag \\
    &~&~~~~~~~~~~~~~~~~~~~~~~~~~+ \frac{x_m^2(\!-\! C_{12}^2 \!+\! 6C_{12} \!-\! 2C_{12} C_{14} \!+\! 2C_{14}
                                                 \!-\! 2C_{24}) \!+\! x_m^4/2}
                              {(x_m^2 \!-\! 2C_{24} \!-\! x_s^2)(x_m^2 \!-\! 2C_{14} \!-\! x_s^2)} \Biggr]
\eea
\bea
I^{e22}_{d\Omega}&=& \sqrt{1-x­_m^2}
\int d\Omega_2 d\Omega_4 \frac{4C_{12} C_{14} \!+\! 2C_{12} C_{24} \!-\! 2C_{14}C_{24} \!-\! 4x_m^2 C_{12} \!+\! 4x_m^2 C_{14}
                               \!+\! x_m^2 C_{24} - x_m^4}
                              {(x_m^2 \!-\! 2C_{24} \!-\! x_s^2)^2}
\eea
\bea
I^{e23}_{d\Omega}&=& \sqrt{1-x­_m^2}
\int d\Omega_2 d\Omega_4 \Biggl[\frac{2C_{12} C_{24}^2 \!-\! 4C_{12}C_{14} \!-\! 4C_{12}C_{24} \!+\! 2C_{12}C_{14}C_{24}}
                                     {(x_m^2 \!-\! 2C_{24} \!-\! x_s^2)(x_m^2 \!-\! 2C_{14} \!-\! x_s^2)} \notag \\
    &~&~~~~~~~~~~~~~~~~~~~~~~~~~+ \frac{x_m^2(\!-\! C_{12}^2 \!+\! 6C_{12} \!-\! 2C_{12} C_{24} \!-\! 2C_{14}
                                                 \!+\! 2C_{24}) \!+\! x_m^4 C_{12}/2}
                              {(x_m^2 \!-\! 2C_{24} \!-\! x_s^2)(x_m^2 \!-\! 2C_{14} \!-\! x_s^2)} \Biggr]
\eea
\bea
I^{e24}_{d\Omega}&=& \!-\! \sqrt{1-x­_m^2}
\int d\Omega_2 d\Omega_4 \Biggl[\frac{2C_{12}^2 C_{14} \!-\! 4C_{12}C_{14} \!+\! 2C_{12}^2 C_{24} \!-\! 4C_{12}C_{24}}
                                     {(x_m^2 \!-\! 2C_{24} \!-\! x_s^2)(x_m^2 \!-\! 2C_{14} \!-\! x_s^2)} \notag \\
    &~&~~~~~~~~~~~~~~~~~~~~~~~~~+ \frac{x_m^2(\!-\! 4C_{12}^2 \!+\! 5C_{12} \!+\! C_{12} C_{14} \!+\! C_{12} C_{24} \!+\! C_{14}                             							\!+\! C_{24})}
                              {(x_m^2 \!-\! 2C_{24} \!-\! x_s^2)(x_m^2 \!-\! 2C_{14} \!-\! x_s^2)} \Biggr]
\eea
\bea
I^{e33}_{d\Omega}&=& \sqrt{1-x­_m^2}
\int d\Omega_2 d\Omega_4 \frac{2C_{12}C_{14} \!+\! 4C_{12}C_{24} \!-\! 2C_{14}C_{24} \!+\!
                               x_m^2(\!-\! 4C_{12} \!+\! C_{14})}
                              {(x_m^2 \!-\! 2C_{14} \!-\! x_s^2)^2}
\eea
\bea
I^{e34}_{d\Omega}&=& \!-\! \sqrt{1-x­_m^2}
\int d\Omega_2 d\Omega_4 \Biggl[\frac{\!-\!2C_{12}C_{14}^2 \!+\! 4C_{12}C_{14} \!+\! 2C_{14}^2C_{24} \!+\!8C_{12}C_{24}
                                      \!-\!4C_{12}C_{14}C_{24} \!-\! 4C_{14}C_{24}}
                                     {(x_m^2 \!-\! 2C_{14} \!-\! x_s^2)^2} \notag \\
    &~&~~~~~~~~~~~~~~~~~~~~~~~~~+ \frac{x_m^2(2C_{12}^2 \!-\! 8C_{12} \!+\! C_{12}C_{14} \!-\! C_{14}C_{24} \!+\!4C_{24})
                                        -x_m^4}
                              {(x_m^2 \!-\! 2C_{14} \!-\! x_s^2)^2} \Biggr]
\eea
\bea
I^{e44}_{d\Omega}&=& \sqrt{1-x­_m^2}
\int d\Omega_2 d\Omega_4 \frac{2C_{12}C_{14}\!+\! 4C_{12}C_{24} \!-\! 2C_{14}C_{24} \!+\!
                               x_m^2(\!-\! 4C_{12} \!+\! C_{14} \!+\! 4C_{24}) - x_m^4}
                              {(x_m^2 \!-\! 2C_{14} \!-\! x_s^2)^2}
\eea

\bea
I^{\mu 11}_{d\Omega} &=& \left(1-x­_m^2\right)^{3/2}
\int d\Omega_1 d\Omega_3 \frac{\!-\!2\barC_{13}\barC_{14} \!+\! 2\barC_{13}\barC_{34} \!+\! 4\barC_{14}\barC_{34}
                               \!+\! x_m^2(3\barC_{14} \!-\! 5\barC_{34}) \!-\! x_m^4}
                              {(2x_m^2 \!-\! 2\barC_{13} \!-\! x_s^2)^2}
\eea
\bea
I^{\mu 12}_{d\Omega}&=& \!-\! \left(1-x­_m^2\right)^{3/2}
\int d\Omega_1 d\Omega_3 \Biggl[\frac{\!-\!2\barC_{13}\barC_{14}^2 \!+\! 4\barC_{14}^2 \!-\!2\barC_{13}\barC_{34}^2
                               \!-\!4\barC_{34}^2 \!+\!8\barC_{13}\barC_{34} \!+\!8\barC_{14}\barC_{34}}
                              {(2x_m^2 \!-\! 2\barC_{13} \!-\! x_s^2)^2} \notag \\
    &~&~~~~~~~~~~~~~~~~~~~~~~~~~+ \frac{x_m^2(4\barC_{14}^2 \!+\! 4\barC_{34}^2 \!-\!8\barC_{13} \!+\!2\barC_{13}\barC_{14}
                                        \!-\!8\barC_{14} \!-\!2\barC_{13}\barC_{34} \!-\!4\barC_{14}\barC_{34}
                                        \!-\!8\barC_{34})}
                              {(2x_m^2 \!-\! 2\barC_{13} \!-\! x_s^2)^2} \notag \\
    &~&~~~~~~~~~~~~~~~~~~~~~~~~~+ \frac{x_m^4(3\barC_{13} \!-\!2\barC_{14} \!+\!2\barC_{34} \!+\! 10) \!-\! 3x_m^6}
                              {(2x_m^2 \!-\! 2\barC_{13} \!-\! x_s^2)^2} \Biggr]
\eea
\bea
I^{\mu 13}_{d\Omega}&=& \!-\! \left(1-x­_m^2\right)^{3/2}
\int d\Omega_1 d\Omega_3 \Biggl[\frac{2\barC_{13}\barC_{34}^2 \!+\! 2\barC_{14}\barC_{34}^2 \!-\! 4\barC_{13}\barC_{34}
                                      \!-\! 4\barC_{14} \barC_{34} }
                              {(2x_m^2 \!-\! 2\barC_{13} \!-\! x_s^2)(2x_m^2 \!-\! 2\barC_{14} \!-\! x_s^2)} \notag \\
    &~&~~~~~~~~~~~~~~~~~~~~~~~~~+ \frac{x_m^2(\!-\!4\barC_{34}^2 \!+\! 5\barC_{13} \!+\! 5\barC_{14} \!+\!\barC_{13}\barC_{34}
                                        \!+\!\barC_{14}\barC_{34} \!+\! 5\barC_{34})}
                              {(2x_m^2 \!-\! 2\barC_{13} \!-\! x_s^2)(2x_m^2 \!-\! 2\barC_{14} \!-\! x_s^2)} \notag \\
    &~&~~~~~~~~~~~~~~~~~~~~~~~~~+ \frac{x_m^4(\!-\!3\barC_{13}/2 \!-\! 3\barC_{14}/2 \!-\! \barC_{34} \!-\!6) \!+\! 2x_m^6}
                              {(2x_m^2 \!-\! 2\barC_{13} \!-\! x_s^2)(2x_m^2 \!-\! 2\barC_{14} \!-\! x_s^2)} \Biggr]
\eea
\bea
I^{\mu 14}_{d\Omega} &=& \left(1-x­_m^2\right)^{3/2}
\int d\Omega_1 d\Omega_3 \Biggl[\frac{2\barC_{13}\barC_{34}^2 \!+\! 4\barC_{34}^2 \!-\! 8\barC_{13}\barC_{34}
                                      \!-\! 8\barC_{14}\barC_{34} }
                              {(2x_m^2 \!-\! 2\barC_{13} \!-\! x_s^2)(2x_m^2 \!-\! 2\barC_{14} \!-\! x_s^2)} \notag \\
    &~&~~~~~~~~~~~~~~~~~~~~~~~~~+ \frac{x_m^2(\!-\!\barC_{14}^2 \!-\! 3\barC_{34}^2 \!+\! 6\barC_{13} \!+\! 8\barC_{14}
                                        \!+\! 2\barC_{13}\barC_{34} \!+\! 2\barC_{14}\barC_{34} \!+\! 8\barC_{34})}
                              {(2x_m^2 \!-\! 2\barC_{13} \!-\! x_s^2)(2x_m^2 \!-\! 2\barC_{14} \!-\! x_s^2)} \notag \\
    &~&~~~~~~~~~~~~~~~~~~~~~~~~~+ \frac{x_m^4(\!-\!3\barC_{13}/2 \!-\! 2\barC_{34} \!-\!9) \!+\! 3x_m^6/2}
                              {(2x_m^2 \!-\! 2\barC_{13} \!-\! x_s^2)(2x_m^2 \!-\! 2\barC_{14} \!-\! x_s^2)} \Biggr]
\eea
\bea
I^{\mu 22}_{d\Omega}&=& \left(1-x­_m^2\right)^{3/2}
\int d\Omega_1 d\Omega_3 \frac{2\barC_{14}^2 \!-\! 2\barC_{34}^2 \!+\! 4\barC_{13}\barC_{34} \!+\! 4\barC_{14}\barC_{34}
                               \!+\!x_m^2(\!-\!4\barC_{13}\!-\!4\barC_{34})\!+\!4x_m^4}
                              {(2x_m^2 \!-\! 2\barC_{13} \!-\! x_s^2)^2}
\eea
\bea
I^{\mu 23}_{d\Omega}&=& \left(1-x­_m^2\right)^{3/2}
\int d\Omega_1 d\Omega_3 \Biggl[\frac{2\barC_{14}\barC_{34}^2 \!+\! 4\barC_{34}^2 \!-\! 8\barC_{13}\barC_{34}
                                      \!-\! 8\barC_{14}\barC_{34} }
                              {(2x_m^2 \!-\! 2\barC_{13} \!-\! x_s^2)(2x_m^2 \!-\! 2\barC_{14} \!-\! x_s^2)} \notag \\
    &~&~~~~~~~~~~~~~~~~~~~~~~~~~+ \frac{x_m^2(\!-\!\barC_{13}^2 \!-\! 3\barC_{34}^2 \!+\! 10\barC_{13} 
                                        \!+\! \barC_{13}\barC_{14}\!+\! 6\barC_{14}
                                        \!+\! 2\barC_{13}\barC_{34} \!+\! 2\barC_{14}\barC_{34} \!+\! 8\barC_{34})}
                              {(2x_m^2 \!-\! 2\barC_{13} \!-\! x_s^2)(2x_m^2 \!-\! 2\barC_{14} \!-\! x_s^2)} \notag \\
    &~&~~~~~~~~~~~~~~~~~~~~~~~~~+ \frac{x_m^4(\!-\!\barC_{13} \!-\! 2\barC_{14} \!-\! 2\barC_{34} \!-\!10) \!+\! 2x_m^6}
                              {(2x_m^2 \!-\! 2\barC_{13} \!-\! x_s^2)(2x_m^2 \!-\! 2\barC_{14} \!-\! x_s^2)} \Biggr]
\eea
\bea
I^{\mu 24}_{d\Omega}&=& \!-\! \left(1-x­_m^2\right)^{3/2}
\int d\Omega_1 d\Omega_3 \frac{4\barC_{34}^2 \!-\! 8\barC_{13}\barC_{34} \!-\! 8\barC_{14}\barC_{34}
                                      \!+\!x_m^2(8\barC_{13} \!+\!8\barC_{14} \!+\! 8\barC_{34}) -8x_m^4}
                              {(2x_m^2 \!-\! 2\barC_{13} \!-\! x_s^2)(2x_m^2 \!-\! 2\barC_{14} \!-\! x_s^2)}    
\eea
\bea
I^{\mu 33}_{d\Omega}&=& \left(1-x­_m^2\right)^{3/2}
\int d\Omega_1 d\Omega_3 \frac{\!-\!2\barC_{14}^2 \!+\! 2\barC_{14}\barC_{34} \!+\! x_m^2(\barC_{13} \!+\!6\barC_{14}
                               \!-\!3\barC_{34})\!-\!3x_m^4}
                              {(2x_m^2 \!-\! 2\barC_{14} \!-\! x_s^2)^2}
\eea
\bea
I^{\mu 34}_{d\Omega}&=& \!-\! \left(1-x­_m^2\right)^{3/2}
\int d\Omega_1 d\Omega_3 \Biggl[\frac{2\barC_{13}^2 \!-\! 2\barC_{13}\barC_{14}^2 \!+\! 2\barC_{13}\barC_{14} 
                                      \!-\!2\barC_{13}\barC_{34} \!+\! 2\barC_{13}\barC_{14}\barC_{34}}
                              {(2x_m^2 \!-\! 2\barC_{14} \!-\! x_s^2)^2} \notag \\
    &~&~~~~~~~~~~~~~~~~~~~~~~~~~+ \frac{x_m^2(\!-\!\barC_{13}^2 \!-\!\barC_{34}^2 \!+\!2\barC_{13} \!+\!4\barC_{13}\barC_{14}
                                        \!+\! 2\barC_{14})}
                              {(2x_m^2 \!-\! 2\barC_{14} \!-\! x_s^2)^2} \notag \\
    &~&~~~~~~~~~~~~~~~~~~~~~~~~~+ \frac{x_m^4(\!-\!2\barC_{13} \!-\!2\barC_{14} \!-\!2) \!+\! 2x_m^6}
                              {(2x_m^2 \!-\! 2\barC_{14} \!-\! x_s^2)^2}\Biggr]
\eea
\bea
I^{\mu 44}_{d\Omega}&=& \left(1-x­_m^2\right)^{3/2}
\int d\Omega_1 d\Omega_3 \frac{2\barC_{13}^2 \!-\! 2\barC_{34}^2 \!+\! 4\barC_{13} \barC_{34} \!+\! 4\barC_{14}\barC_{34}
                               \!+\! x_m^2(\!-\!4\barC_{14} \!-\!4\barC_{34})\!+\!4x_m^4}
                              {(2x_m^2 \!-\! 2\barC_{14} \!-\! x_s^2)^2}
\eea

\renewcommand{\href}[2]{#2}

\newcommand{\apjl}{Astrophys. J. Lett.\ }
\newcommand{\mnras}{Mon. Not. R. Astron. Soc.\ }
\newcommand{\aap}{Astron. Astrophys.\ }

\bibliographystyle{JHEP_MGA}
\bibliography{leptonic_viscosity} 

\end{document}